\documentclass[prd,aps,twocolumn,preprintnumbers,amsmath,amssymb,nofootinbib,superscriptaddress,notitlepage]{revtex4-1}

\usepackage{epsfig}
\usepackage[utf8]{inputenc}
\usepackage{color}
\usepackage{epstopdf}
\usepackage[colorlinks=true,
linkcolor=blue,
breaklinks=true,
urlcolor=blue,
citecolor=blue]{hyperref}

\newcommand{\nn}{\nonumber}
\newcommand{\be}{\begin{equation}}
\newcommand{\ee}{\end{equation}}
\newcommand{\bea}{\begin{eqnarray}}
\newcommand{\eea}{\end{eqnarray}}
\newcommand{\ba}{\begin{array}}
\newcommand{\ea}{\end{array}}
\newcommand{\bi}{\begin{itemize}}
\newcommand{\ei}{\end{itemize}}

\newcommand{\mcb}{{\mathcal B}}

\newcommand{\lf}{\left}
\newcommand{\rg}{\right}

\newcommand{\blue}[1]{{\color{blue}#1}}

\newcommand{\ucas}{\affiliation{University of Chinese Academy of Sciences, Beijing 100049, China}}

\newcommand{\imp}{\affiliation{Institute of Modern Physics, Chinese Academy of Sciences, Lanzhou 730000, China}}

\newcommand{\csr}{\affiliation{Research Center for Hadron and CSR Physics, Lanzhou University and Institute of Modern Physics of CAS, Lanzhou 730000, China}}

\newcommand{\giessen}{\affiliation{Institut f\"{u}r Theoretische Physik, Universit\"{a}t Giessen, D-35392 Giessen, Germany}}

\begin{document}

\title{Data driven isospin analysis of timelike octet baryons electromagnetic form factors and charmonium decay into baryon-anti-baryon}

\author{Jian-Ping Dai}\email{daijianping@ynu.edu.cn}
\affiliation{Department of Physics, Yunnan University, Kunming 650091, China}

\author{Xu Cao}\email{caoxu@impcas.ac.cn}
\imp
\ucas
\csr

\author{Horst Lenske}\email{horst.lenske@physik.uni-giessen.de}
\giessen

\date{\today}

\begin{abstract}
  \rule{0ex}{3ex}
Inspired by the recent precise data,
we perform model-independently an isospin decomposition of the timelike octet baryons electromagnetic form factors.
As noted in our previous work, the relative magnitude of isoscalar and isovector component is determined with the input of data on various isospin channels.
Herein we further assert that {their relative phase can be constraint by the phase difference of oscillatory modulation of effective form factors} between isospin channels.
The framework is extended to analyze the data of differential cross sections and applied to the form factors of nucleon and hyperons with detail and isospin non-conservation of charmonium decay into baryon-anti-baryon as well. We address that isospin analysis is meaningful when the isospin broken scale is compared to or smaller than the uncertainties of data.

\end{abstract}

\maketitle

\section{Introduction} \label{sec:intro}

As the lightest baryons, nucleons account essentially for all of the observable matter in the universe. However, since their non-pointlike nature was discovered more than sixty years ago~\cite{Hofstadter:1956qs}, nucleon properties, e.g. the spin and radii, are still a puzzle, whose final solution is pending~\cite{Ji:2020ena,Pohl:2010zza,Bezginov:2019mdi,Xiong:2019umf}.
Moreover, understanding the structure of nucleons is of high importance for the physics of the whole set of baryons forming the  lowest SU(3) octet. On an astrophysical scale, hyperons with  a strange quark  content, are expected to play a key role  to understand the masses and sizes of compact object like neutron stars \cite{Bombaci:2016xzl}.
Electromagnetic form factors (EMFFs) are fundamental observables of baryons that are closely related to their static and dynamical properties~\cite{Denig:2012by,Pacetti:2014jai}. The transition amplitude of electron-positron annihilation into baryon-anti-baryon (or \textit{vice versa}) 
is given by timelike EMFFs, which are complex and connected with the spacelike EMFFs by the dispersion relation~\cite{Belushkin:2006qa}. The known sources of the imaginary part of amplitudes are multi-meson rescattering loops and intermediate vector meson excitations. While the former are incorporated by the dispersion analysis of the nucleon EMFFs including meson continua~\cite{Belushkin:2006qa,Lin:2021xrc,Lin:2022baj}, the latter are only phenomenologically investigated in the isoscalar and isovector form factor~\cite{Iachello:1972nu}.

In view of the precise data measured by several collaborations ~\cite{Huang:2021xte,Xia:2021agf,Lin:2022hyr,Larin:2022jxp}, it is timely to perform a data driven analysis based on the isospin decomposition.
Though isospin conservation is given only approximately, the analysis is model-independent by its own provided that isospin mixing is allowed.
Isospin decomposition is in fact far from trivial when isoscalar and isovector components are both involved into $e^+e^- \to B \bar B$ reactions. At first glimpse, if one measures all the isospin channels, the isospin assignment of contributing vector meson above $B\bar B$ threshold shall be unambiguously determined. However, the isospin assignment of a light-quark meson or charmonium state is in fact highly non-trivial 
when both isoscalar and isovector components are populated in $e^+ e^-$ reactions. 
Take $N \bar N$ production as an example, two complex isospin amplitudes with at least three parameters (two moduli and one relative phase) are present but only two channels ($p \bar p$ and $n \bar n$) of mixed isospin ($I$=0, 1) are allowed. As a result, one can only constrain the range of two parameters, namely relative magnitude and relative phase if total cross sections (or equivalently effective form factors) are measured.
It is further noted that complex electric and magnetic form factors are involved in the analysis of angular distribution and polarization observables. As a result, one cannot fully decompose all complex amplitudes, even if all observables were measured.

Following exactly the same argument, the isospin assignment of involved vector mesons cannot be determined in principle. Note that experimentally one knows the isospin of $J/\psi$ and other $1^{--}$ pieces of charmonium not from their production in $e^+ e^-$ annihilation, but from their strongly suppressed isovector decays and the absence of charged partners. This is totally different from the case of $\gamma N$ or $\pi N \to \pi N$ reactions \cite{Cao:2013psa}, where the full set of isospin channels is available to disentangle explicitly two isospin amplitudes $I = 1/2$ and $3/2$. A good example is $\rho(2150)$ and $\phi(2175)$, whose masses and widths with big uncertainties are overlapping. When only one of them is contributing and another one is forbidden by isospin conservation, the issue on isospin is obvious by itself. For instance, only $\rho(2150)$ is present in $K^+K^-$ spectrum of $e^+ e^- \to J/\psi\to K^+K^-\pi^0$~\cite{BESIII:2019apb}, while only $\phi(2175)$ in $e^+ e^- \to \phi \pi^+ \pi^-$~\cite{BESIII:2021lho}. However, the broad peak around 2.0 $\sim$ 2.2 GeV clearly observed both in $e^+ e^- \to K^+ K^-$ \cite{BESIII:2018ldc} and $e^+ e^- \to K^+ K^- \pi^0 \pi^0$ \cite{BESIII:2020vtu}, would be from $\rho(2150)$ or $\phi(2175)$ or both, because the isospin of this peak is never determined. The $\omega$(2290) with higher mass, solely contributing to $e^+ e^- \to \omega \pi^0 \pi^0$ \cite{BESIII:2021uni}, is also possibly contributing through interference considering its broad width. Under these circumstances,  both $\rho(2150)$ and $\phi(2175)$ are probably contributing into these channels and their isospin cannot unambiguously be disentangled by their own.

In a word, isospin assignment of the involved vector mesons in $e^+ e^- \to B \bar B$ cannot be determined from the $B \bar B$ data alone, except the special case of a $\Lambda$ hyperon. More uncomfortably, in comparison with $K^+K^-$ and $K^+K^-\pi^0 \pi^0$ channels, the broad structure of $N \bar N$ spectra shows a global oscillation pattern~\cite{Bianconi:2015owa,Bianconi:2015vva,Tomasi-Gustafsson:2020vae} rather than resonant Breit-Wigner structures which resulted into the long confusion about their underlying mechanism. 
The proposed interpretations ranged from effects induced by the nucleon internal structure~\cite{Bianconi:2015owa,Tomasi-Gustafsson:2022tpu}, vector meson excitations \cite{deMelo:2008rj,deMelo:2005cy,Lorenz:2015pba}, threshold opening~\cite{Lorenz:2015pba}, or final state interactions \cite{Yang:2022qoy,Qian:2022xib}.
In our previous paper we clarifed the case on nucleon by introducing an isospin formalism allowing to decompose the amplitudes and cross sections \cite{Cao:2021asd}, thus paving the way to a data driven isospin analysis of the continuum component containing nucleon structure in timelike region as detailed in Sec. \ref{sec:nucleon}.
An extension to octet baryons is obvious and will be done in Sec. \ref{sec:hyperon}.
The whole framework is easily extended to analyze the isospin non-conservation of charmonium decay into baryon-anti-baryon in Sec. \ref{sec:charm}.

\section{About nucleons} \label{sec:nucleon}

{The basics on the observables of $B \bar{B}$ production are shown in the Appendix \ref{apx:polar}, following the usual definition in the literature.}
The effective nucleon form factor $|G^N_{\textrm{eff}}|$ shows an oscillating behavior around a smooth component $G_N^{D} (q^2)$,
\bea\label{eq:Neff}
   |G^N_{\textrm{eff}}| &=&  |G_{N}^D(q^2)| + G_N^{\textrm{osc}}(q^2),
\\ |G_N^D(q^2)| &=& \frac{\mathcal{A}_{N}}{(1+\frac{q^2}{m_a^2})(1-\frac{q^2}{0.71 ( \textrm{GeV}^2)})^2}, \label{eq:dipole}
\eea
{which are fitted well to the data of interest but do not satisfy the pertubative QCD expectation for large $q^2$.
These ambiguities reflect the persisting lack of a first principles approach to a desired precision, awaiting to be solved in the future hopefully by LQCD \cite{Alexandrou:2018sjm}, once physical parameters are used, or by an effective field theory \cite{Mondal:2019jdg}.
However, conclusive answers from low to high-$q^2$ are still pending because of their complexity.
Obviously any other proper parametrizations work equally well in the later isospin analysis driven by data.
}
For protons ($N=p$), the dipole parameters for the normalization factor $\mathcal{A}_p$ and the pole mass $m^2_a$ have been determined from the fit to the $|G_p|$ results from the BaBar experiment as $\mathcal{A}_p = 7.7$ and $m^2_a = 14.8$ GeV$^2$.
The normalization for the neutron was obtained to be $\mathcal{A}_n = 4.87 \pm 0.09$ with the {same pole mass}
as for the proton EFF~\cite{BESIII:2021dfy}.

The oscillatory structure $G_N^{\textrm{osc}}$ is parameterized as \cite{Bianconi:2015owa},
\bea \label{eq:osc}
G_N^{\textrm{osc}} &=& A_{N} \textrm{exp}\left(-B_N \, p \right)\, \cos \left(C_N \, p + \it{D_N}\right),
\eea
with $p = \sqrt{q^2(q^2/4m^2_N -1)}$ being the nucleon momentum in the antinucleon rest frame.
The normalization $A_{N} \ll \mathcal{A}_{N}$ is anticipated because no wide vector meson is found to couple strongly to $N\bar{N}$ below the charmonium region.
However, a prominent hadronic counterexample is $e^+ e^- \to \pi^+ \pi^-$, where wide vector mesons, e.g. the $\rho$-meson, are strongly emergent.
In order to keep $G_N^{\textrm{osc}}/G_N^{D}$ small over a wide energy range the damping strength shall be strong enough, e.g. $B_N \gtrsim 0.5$ GeV$^{-1}$.
The momentum frequency shall be in a moderate range, $2 \lesssim C_N \lesssim 10$ GeV$^{-1}$.
Otherwise it will over-fit the data with artificial narrow structures when frequency is chosen too small ($C_N > 10$ GeV$^{-1}$),
or it contains too wide continuum which can be absorbed into the smooth component when the frequency is too large ($C_N < 2$ GeV$^{-1}$).
The fit to proton data in various schemes does follow these expectations \cite{Tomasi-Gustafsson:2020vae,BESIII:2021dfy}.
The neutron data of DM2~\cite{Biagini:1990nb}, FENICE~\cite{Antonelli:1998fv}, however, suffer from a limited energy range and low statistics.
The near threshold SND data \cite{Achasov:2014ncd,Druzhinin:2019gpo,SND:2022wdb} are more involved the final state interaction, which need to be careful considered~\cite{Yang:2022qoy,Qian:2022xib}.
The residual component is found to be $A_p = 0.07 \pm 0.01$ and $A_n = 0.08 \pm 0.03$ with universal parameters for proton and neutron \cite{BESIII:2021dfy}: $B_N = 1.01 \pm0.24 $ GeV$^{-1}$, $C_N = 5.28\pm 0.36$  GeV$^{-1}$.
The phases are $D_p =$ 0.31 $\pm$ 0.17 and $D_n$ = -3.77 $\pm$ 0.55, resulting into a relative phase for the proton and neutron oscillation being $\Delta \phi = (125 \pm 12)^\circ$.

\begin{figure}[t]
    \centering
    \includegraphics[width=\linewidth]{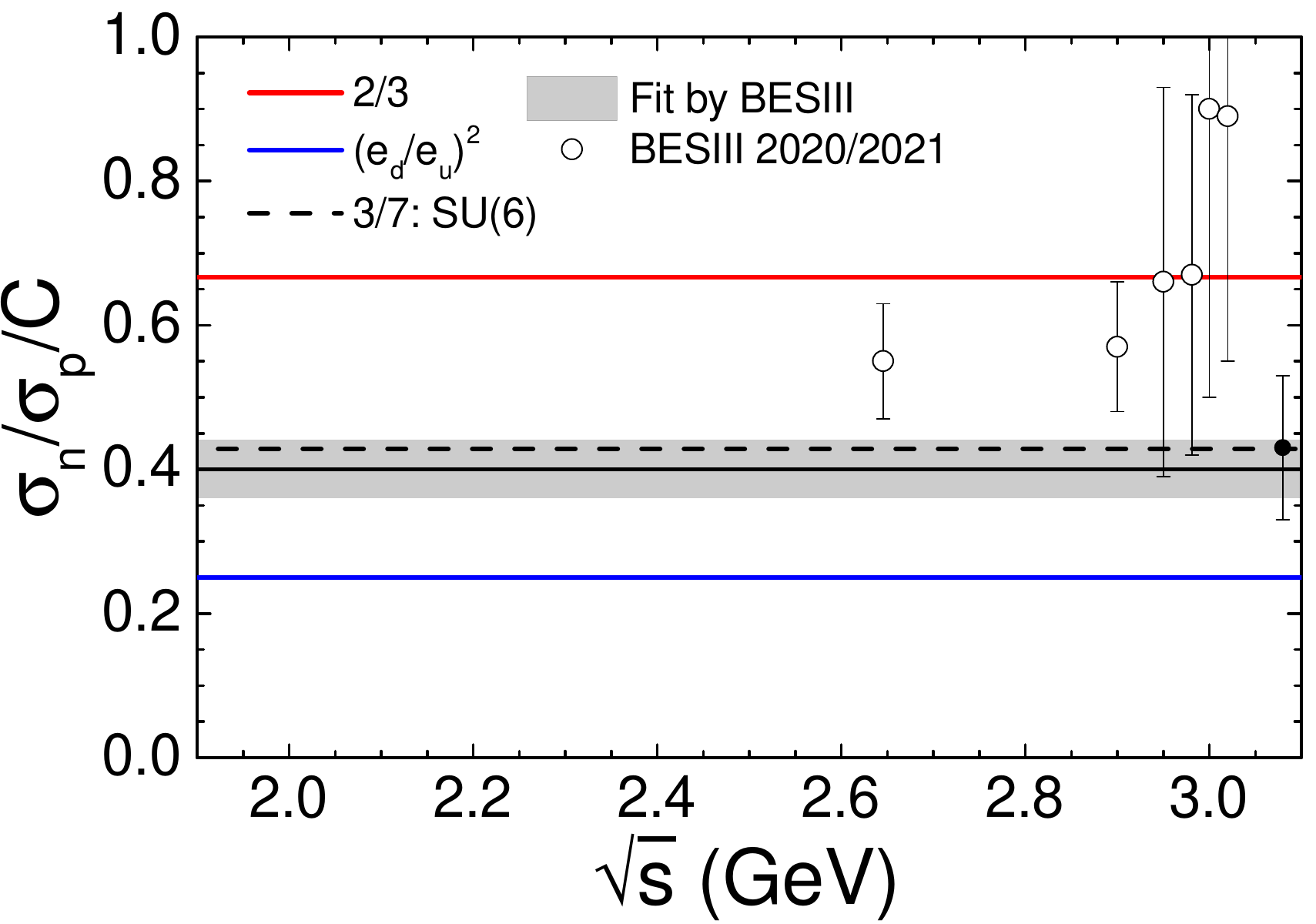}
    \caption{(Color online) The cross section ratio of neutron to proton after subtracting the oscillation component and Coulomb correction. The BESIII data above 2.6 GeV are shown for comparison ~\cite{BESIII:2021dfy}.}
    \label{fig:rationp}
\end{figure}

As argued in our previous work \cite{Cao:2021asd},
such oscillatory structures are the natural consequence of an interference between a leading component from nucleon structure and a residual amplitude from vector meson production above threshold.
Thus, a ratio containing no oscillation is suggested,
\be \label{eq:rationpn}
\frac{\sigma^D_n}{\sigma^0_p/C} =\left |\frac{G^D_n}{G^D_p} \right |^2 = 0.40 \pm 0.03
\ee
surprisingly a constant, albeit within uncertainties, over a wide range of energies.
It shall be noted that above errors do not include those from different parameterization scheme of the smooth component in Eq. (\ref{eq:dipole}). Particularly its robustness shall be tested by the future neutron data with better precision, as have done in the proton case~\cite{Tomasi-Gustafsson:2020vae}.
In Fig. \ref{fig:rationp}, this constant is compared to several values predicted for the spacelike region,
namely the 3/7, a prediction of SU(6) symmetric nucleon wave function~\cite{Farrar:1975yb},
together with the naive $e_u^2/e_d^2 = $1/4 from the quark charge ratio~\cite{Chernyak:1984bm} and 2/3 obtained from the values in constituent quark model~\cite{Farrar:1975yb}.
It is also consistent with 0.4$\sim$0.5 predicted by QCD sum rule \cite{Chernyak:1984bm} and $(\mu_n/\mu_p)^2$ deduced from the asymptotic behaviour \cite{Bloom:1970xb}.
A recent theoretical analysis of the MARATHON
experiment~\cite{MARATHON:2021vqu} finds quite-close values~\cite{Segarra:2019gbp,Cui:2021gzg,Chang:2022jri}.

\begin{figure}[t]
    \centering
    \includegraphics[width=\linewidth]{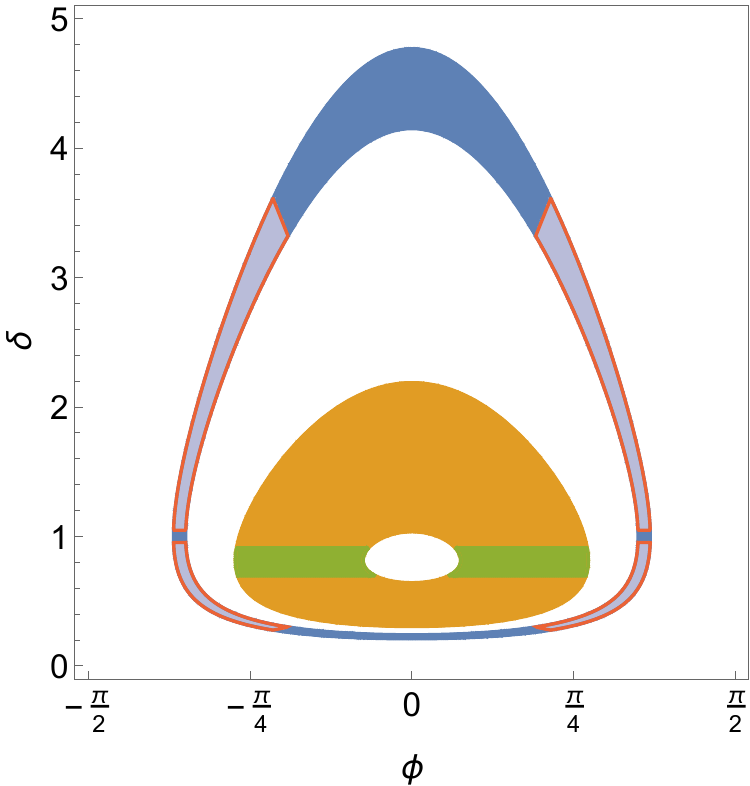}
    \caption{(Color online) Permeable boundaries of relative magnitude $\delta$ and phase $\phi$ (see Eqs. (\ref{eq:dataRID}, \ref{eq:dataphi}) for nucleon (outer region) and $\Sigma$ (inner region), respectively. }
    \label{fig:GSosc_bound}
\end{figure}

For a long time, there was barely any information on vector mesons above 2.6 GeV \cite{Wang:2021abg,Liu:2022yrt}.
In Fig. \ref{fig:rationp}, the deviation of BESIII data from the constant in this energy range ~\cite{BESIII:2021dfy} shall be interpreted as the possible resonances with larger mass.
The bold point just below the $J/\psi$ mass peak is not supposed to become contaminated with light flavor vector mesons.
Within the large error bars it is actually consistent with our postulation.
This further supports our argument that the oscillation part $G_N^{osc}$ is probably connected to vector meson excitation while the leading component $G_{N}^D$ is what is actually related directly to the nucleon structure.
In terms of the  $|I, I_3 \rangle$ = $|0, 0 \rangle$ and $|1, 0 \rangle$ isospin components,
this leading part is decomposed into the isoscalar $I_0$ and isovector $I_1$ amplitudes with regard to isospin symmetry \cite{Ellis:2001xc}:
\be \label{eq:isoamp}
G^D_{p,n} = \frac{I_1^D\pm I_0^D}{\sqrt{2}} \,,
\ee
where the upper (lower) sign corresponds to the proton (neutron) case.
The two isospin amplitudes are related by
$I_1^D = I_0^D \delta_I \, e^{i \phi_I} $, rendering an isospin relevant ratio \cite{Cao:2021asd}:
\be \label{eq:dataRID}
R_I^{\textrm{eff}} = \frac{|G^D_p|^2 - |G^D_n|^2}{|G^D_p|^2 + |G^D_n|^2} =\frac{2\delta_I \cos{\phi_I}}{1 + \delta_I^2}  = 0.43 \pm 0.03,
\ee
{whose significant deviation from zero is a clear signature of its difference from resonant states decay.} It constrains the range of $\delta_I$ and $\phi_I$ within an blue egg-shaped diagram as shown in Fig. \ref{fig:GSosc_bound}.
Considering the smallness of $G_N^{osc}/|G^D_{N}|\ll 1$~\cite{Cao:2021asd},
\bea
G_N^{\textrm{osc}} &\simeq& \frac{|I^{\textrm{BW}}_N|}{\sqrt{2}}\cos(\phi_N^D-\phi_N^{\textrm{BW}}) \nn \\
&+& \frac{1}{4} \frac{|I^{\textrm{BW}}_N|^2}{G_N^D} \sin^2 (\phi_N^D-\phi_N^{\textrm{BW}})
\label{eq:pexpansion}
\eea
is an equivalent of Eq. (\ref{eq:osc}) regardless of next-to-leading order (NLO) contribution in the second line.
Moreover, another significant result of our present work  was to identify that {at leading order} the phase shift $\phi_p^{\textrm{BW}} = \phi_n^{\textrm{BW}}$ of Breit–Wigner (BW) distribution serves as a measure of the relative phase for the proton and neutron oscillation:
\be \label{eq:dataphi}
|\Delta \phi| = |D_n-D_p| = \arg\frac{I_1^D-I_0^D}{I_1^D+I_0^D},
\ee
thus constraining further the compatible values of $\delta_I$ and $\phi_I$ to rather well defined regions as indicated in the red correlation diagram displayed in Fig. \ref{fig:GSosc_bound}.
Interestingly, both $|I_0| = |I_1|$ and the in-phase scenario $\phi_I = 0$ are excluded with high statistical significance.
{Note that this simple relation receives correction from higher order terms above,
which certainly give rise to more complicated formula of energy dependence.
The precision of measurements up to now is far from being adequate to probe these tiny terms.
}

Assuming the isospin (or charge) symmetry $G^{u/n}_{E,M} = G^{d/p}_{E,M}$ and $G^{d/n}_{E,M} = G^{u/p}_{E,M}$,
the isospin relation is a realization of the flavor decomposition of nucleon EMFFs~\cite{Alexandrou:2018sjm},
\bea \label{eq:isoEMp}
G^p_{E,M} &=& \frac{2}{3} G^u_{E,M} - \frac{1}{3} G^d_{E,M} = \frac{1}{2} \lf(\frac{G^{u+d}_{E,M}}{3} + G^{u-d}_{E,M} \rg) \quad \\ \label{eq:isoEMn}
G^n_{E,M} &=& \frac{2}{3} G^d_{E,M} - \frac{1}{3} G^u_{E,M} = \frac{1}{2} \lf(\frac{G^{u+d}_{E,M}}{3} - G^{u-d}_{E,M} \rg) \quad
\eea
which can be extracted via:
\bea \label{eq:GMsum}
|\frac{G^{u+d}_{M}}{3}|^2 + |G^{u-d}_{M}|^2 &=& 2 (2\tau +1) \lf( \frac{|G_p^D|^2}{2\tau + R_p^2} + \frac{|G_n^D|^2}{2\tau + R_n^2} \rg) \qquad \\ \label{eq:GMsub}
\Re [ \frac{G^{u + d}_M}{3}  G^{u - d \dag}_M ] &=& (2\tau +1) \lf( \frac{|G_p^D|^2}{2\tau + R_p^2} - \frac{|G_n^D|^2}{2\tau + R_n^2} \rg) \quad \\ \label{eq:GEsum}
|\frac{G^{u+d}_{E}}{3}|^2 + |G^{u-d}_{E}|^2 &=& 2 (2\tau +1) \lf( \frac{R_p^2 |G_p^D|^2}{2\tau + R_p^2} + \frac{R_n^2 |G_n^D|^2}{2\tau + R_n^2} \rg) \qquad \\ \label{eq:GEsub}
\Re [ \frac{G^{u + d}_E}{3}  G^{u - d \dag}_E ] &=& (2\tau +1) \lf( \frac{R_p^2 |G_p^D|^2}{2\tau + R_p^2} - \frac{R_n^2 |G_n^D|^2}{2\tau + R_n^2} \rg) \quad
\eea
if EFF $G^D_{N}$ and the electromagnetic form factor ratio $R_{N}$ {(see Eq. (\ref{eq:RB}) for the definition)} are both measured.
{Therefore one cannot fully separate all EMFFs in the isospin basis if only the data of total and differential cross sections are available.
The appendix \ref{apx:polar} shows that this separation is attainable after including further two single-spin observables sensitive to $\Im (G_M G_E^*)$ and $\Re (G_M G_E^*)$, respectively. }

\begin{figure}[t]
    \centering
    \includegraphics[width=\linewidth]{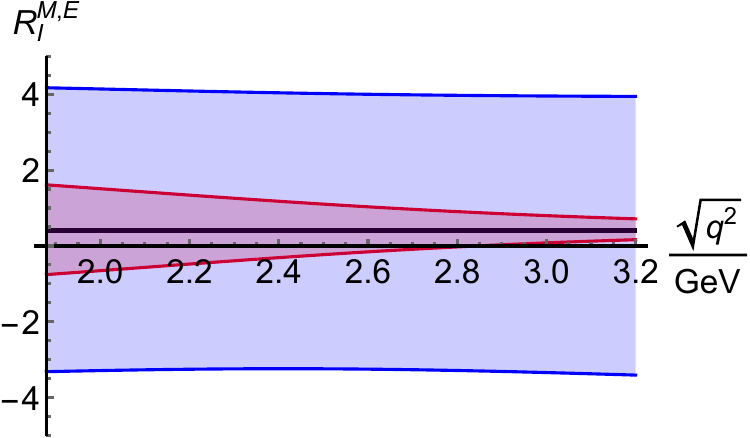}
    \caption{(Color online) The {isospin relevant quantities $R_I^{M}$ in Eq. (\ref{eq:ratioM}) (red region) and $R_I^{E}$ in Eq. (\ref{eq:ratioE}) }(blue region) extracted from data. When $R_p = R_n$ they both return back to the black line, {representing the value in Eq.~(\ref{eq:dataRID})}.}
    \label{fig:interEM}
\end{figure}

{The Eqs. (\ref{eq:GMsum},\ref{eq:GMsub},\ref{eq:GEsum},\ref{eq:GEsub}) are applicable to the whole amplitudes containing both oscillatory and smooth component.
Since the former is attributed dynamically to the vector mesons,
here and also in Sec. \ref{sec:hyperon} we discuss condition involving only the later.
}
Without loss of generality one may use
\be
\frac{G^{u+d}_{M}}{3} = G^{u - d}_{M} \delta_M \, e^{i \phi_M} \,, \quad
\frac{G^{u+d}_{E}}{3} = G^{u - d}_{E} \delta_E \, e^{i \phi_E}.
\ee
With a parameterization of $R_N$ {containing no oscillatory component~\cite{Tomasi-Gustafsson:2020vae} }:
\be \label{eq:RN}
R_N = \frac{r_N}{r_N + (\sqrt{q^2} - 2 m_N)^2}
\ee
one finds $r_p = 3 \pm 2$ GeV$^2$ and $r_n = 6 \pm 21$ GeV$^2$ extracted from the proton data~\cite{Tomasi-Gustafsson:2020vae} and the neutron data~\cite{BESIII:2022rrg}, respectively.
{As a generalization of Eq. (\ref{eq:dataRID})} we could derive the following quantities from Eqs. (\ref{eq:GMsum},\ref{eq:GMsub}) and Eqs. (\ref{eq:GEsum},\ref{eq:GEsub}), respectively
\bea \label{eq:ratioM}
R_I^{M} &=& \frac{|G^p_M|^2 - |G^n_M|^2}{|G^p_M|^2 + |G^n_M|^2} = \frac{2\delta_M \cos{\phi_M}}{1 + \delta_M^2}, \\ 
R_I^{E} &=& \frac{|G^p_E|^2 - |G^n_E|^2}{|G^p_E|^2 + |G^n_E|^2} = \frac{2\delta_E \cos{\phi_E}}{1 + \delta_E^2} \label{eq:ratioE}
\eea
as a function of $\sqrt{q^2}$ in Fig. \ref{fig:interEM}.
If $R_p = R_n$, these ratios are identical to Eq.~(\ref{eq:dataRID}) as shown by the black line in Fig. \ref{fig:interEM}.
Thus, the interference between the isoscalar and isovector EMFFs is observed for the first time, though suffering from big uncertainties.

\section{About hyperons} \label{sec:hyperon}

Many models constructed for nucleons in the timelike region have been extended to
the whole octet for investigations of hyperon EMFFs.
The region close to the reaction threshold is especially interesting
considering the role of final-state interaction, as clearly indicated by an effective optical potential \cite{Milstein:2022tfx}.
A high quality description of both proton and neutron EMFFs has been achieved by utilizing $N \bar{N}$ potential models that have been constructed within chiral effective field theory \cite{Yang:2022qoy}, phenomenological J\"ulich potential \cite{Haidenbauer:2014kja}, and dispersion-theoretical analysis \cite{Lin:2021xrc}.
An extension to hyperon in the timelike region is successful within the same frameworks~\cite{Haidenbauer:2020wyp,Lin:2022baj}.
The properties of these form factors are significantly influenced by the  final state interaction even at higher energies \cite{Qian:2022xib}.
The hyperon EMFFs are also investigated within a covariant quark model~\cite{Ramalho:2019koj,Ramalho:2020laj} and vector meson dominant (VMD) model~\cite{Li:2021lvs,Li:2020lsb,Yan:2023yff}.
A unified view is achieved within big uncertainties if comparing the timelike result to those in spacelike region calculated by chiral perturbation theory \cite{Kubis:2000aa}, light-cone sum rule\cite{Liu:2008yg,Liu:2009mb}, basis light-front quantization~\cite{Xu:2021wwj,Mondal:2019jdg,Peng:2022lte}, and Lattice QCD \cite{Lin:2008mr}.
The role of sea quark contributions to the EMFFs of $\Sigma$ hyperons is addressed within nonlocal chiral effective theory~\cite{Yang:2021odh}.

\begin{figure}[tb]
    \centering
    \includegraphics[width=\linewidth]{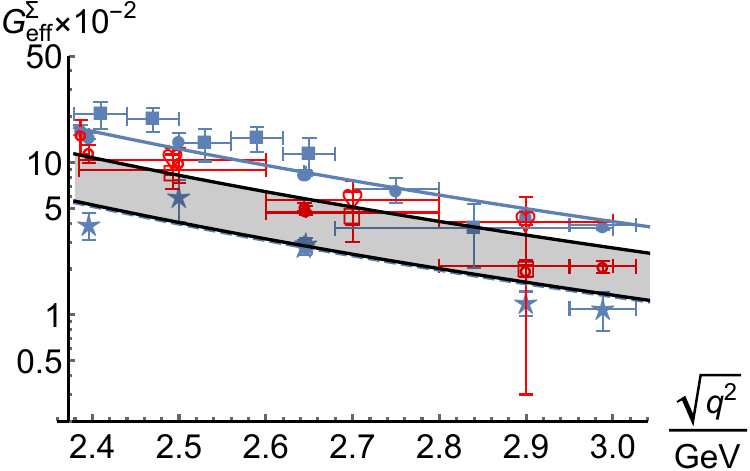}
    \includegraphics[width=\linewidth]{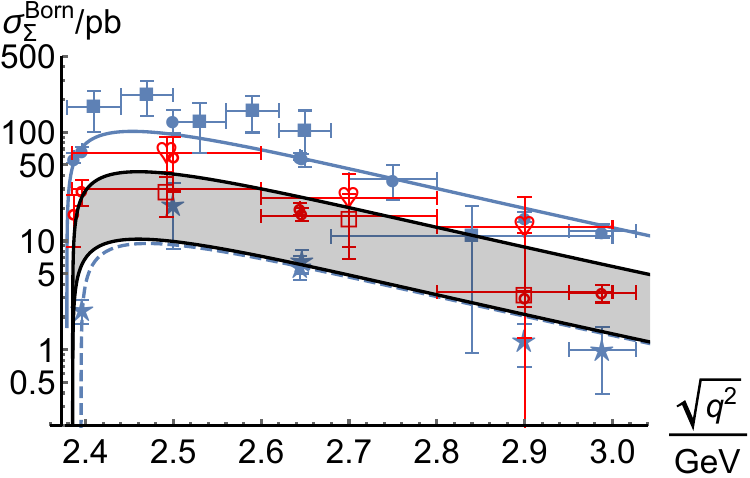}
    \caption{(Color online) The EFF (upper panel) and cross section (lower panel) of $\Sigma^\pm$ and $\Sigma^0$. The dashed and solid line in blue are for $\Sigma^-$ and $\Sigma^+$, respectively. The solid star points in blue are data of $\Sigma^-$ from BESIII \cite{BESIII:2020uqk}. The solid points are data of $\Sigma^+$ from Belle (blue square) \cite{Belle:2022dvb} and BESIII (blue dot) \cite{BESIII:2020uqk}.
    The open red points are data of $\Sigma^0$ from BaBar (square) \cite{BaBar:2007fsu}, Belle (heart) \cite{Belle:2022dvb} and BESIII (dot) \cite{BESIII:2021rkn}.
    The grey bands {lying over the blue dashed lines} are the predicted constraint of $\Sigma^0$ by Eq. (\ref{eq:sigmabound}).
    }
    \label{fig:EFFsigma}
\end{figure}

\begin{figure}
    \centering
    \includegraphics[width=\linewidth]{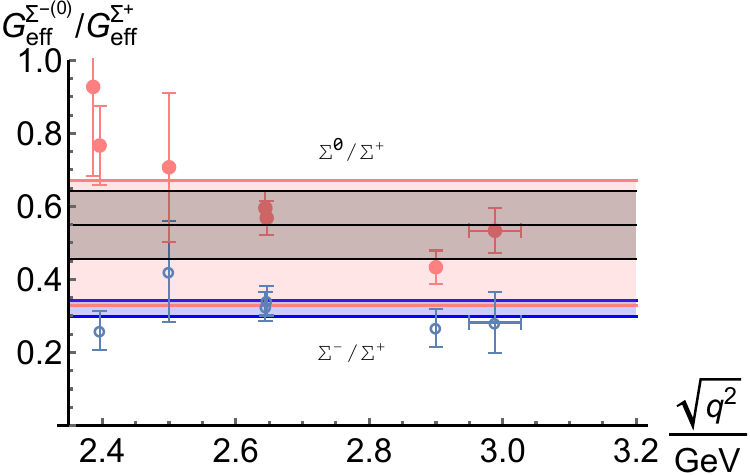}
    \caption{(Color online) The EFF ratio of $\Sigma^-$ to $\Sigma^+$ (blue band and points) is compared to that of $\Sigma^0$ to $\Sigma^+$ (red {and grey bands and red} points). The blue and grey bands are fitted to {the corresponding} data by constants.
    The red band is the predicted constraint by Eq. (\ref{eq:sigmabound}).
    The data are from BESIII \cite{BESIII:2020uqk,BESIII:2021rkn}. {The border of all bands are indicated by the same color of inner region. The central value of grey band is shown by a black line as well.}
    }
    \label{fig:EFFratio}
\end{figure}

\paragraph{Hyperons with Strangeness S=-1}
The striking oscillation phenomenon has not been confirmed for hyperons mainly limited by the data accuracy at hand,
though attempts were made in literature~\cite{Dai:2021yqr,Bianconi:2022yjq}.
However, the precision of data is sufficient for an isospin analysis in line with the spirit in Sec. \ref{sec:nucleon} and also of U-spin considerations~\cite{Biagini:1990nb,Baldini:2007qg}.
The $\Sigma\bar{\Sigma}$ system is of three isospin components $I_0$, $I_1$, and $I_2$, but the $e^+e^- \to \Sigma \bar\Sigma$ reaction populates only the isoscalar and isovector amplitudes $I_0$ and $I_1$.
The corresponding isospin decomposition of $\Sigma^+$, $\Sigma^-$ and $\Sigma^0$ is \cite{Alarcon:2017asr}
\bea \label{eq:isoSigp}
G_+ &=& \frac{1}{\sqrt{2}} I_1 + \frac{1}{\sqrt{3}} I_0 \\  \label{eq:isoSigm}
G_- &=& \frac{1}{\sqrt{2}} I_1 - \frac{1}{\sqrt{3}} I_0 \\  \label{eq:isoSig0}
G_0 &=& \frac{1}{\sqrt{3}} I_0
\eea
Hence if $G_{\pm,0}$ are measured,
both isospin amplitudes can be determined in principle from the sum and the difference where the parameterizations $G_{\pm,0} = |G_{\pm,0}| e^{i \phi_{\pm,0}}$ are to be used. 
Neglecting the small mass difference within the $\Sigma$ isospin triplet, a bound for the EMFF of the neutral channel is deduced:
\be
4\,|G_0|^2 
= |G_+|^2 + |G_-|^2 - 2\, |G_+| |G_-| \cos( \phi_+ - \phi_-) \, \quad
\ee
leading to a model independent bound in a compact form:
\be \label{eq:sigmabound}
||G_+| - |G_-|| \leq 2 |G_0| \leq |G_+| + |G_-|
\ee
This puts forward a constraint to $\Sigma^0$ EFF (and therefore cross section), see {the grey bands} in Fig.~\ref{fig:EFFsigma}.
{Here we use the parameterization of EFF from BESIII \cite{BESIII:2020uqk,BESIII:2021rkn} instead of a {modified} dipole form in Eq.~(\ref{eq:dipole}):}
\be
|G^{\Sigma}_{\textrm{eff}}| = \frac{c_0^{\Sigma}}{(s-c_1)^2(\pi^{2}+\ln^2(s/\Lambda^2_{QCD}))}
\ee
with $c_1 = 1.95$, $c_0^{\Sigma^-} = 0.64$, and $c_0^{\Sigma^+} = 2.0$.
Above bound is surprisingly narrower than the errors of BaBar \cite{BaBar:2007fsu} and Belle \cite{Belle:2022dvb}.
As seen in {the grey bands of} Fig.~\ref{fig:EFFsigma}, the bound of Eq. (\ref{eq:sigmabound}) complies surprisingly well with the  BESIII data \cite{BESIII:2021rkn}.
The good agreement is even clearer in the plot of the EFF ratios of $\Sigma^-$ to $\Sigma^+$ and $\Sigma^0$ to $\Sigma^+$ as shown in Fig.~\ref{fig:EFFratio}.
The error of the predicted ratio between $\Sigma^0$ to $\Sigma^+$ (red band {enclosed by red line}) is a bit worse than those extracted directly from data (grey band {with black border}).

On the other hand, the same procedure as in Sec. \ref{sec:nucleon} could be applied to extract the $I_{0,1}$ amplitudes from the EFFs of the charged members of the $\Sigma$ isospin triplet.
The cross section ratio of $\Sigma^+$ to $\Sigma^-$ is extracted to be $9.7 \pm 1.3$ from recently BESIII data \cite{BESIII:2020uqk}, translated to the blue band in Fig.~\ref{fig:EFFratio}.
This results into
\be \label{eq:dataRIS}
R_I^{\textrm{eff}} = \frac{|G_+|^2 - |G_-|^2}{|G_+|^2 + |G_-|^2} =\frac{2 \sqrt{6} \delta_I \cos{\phi_I}}{3 \delta_I^2 + 2}  = 0.81 \pm 0.16
\ee
which implies a larger interference term between $I_1$ and $I_0$ than that found for the nucleon doublet in Eq. (\ref{eq:dataRID}).
The bound of Eq. (\ref{eq:dataRIS}) is plotted as the inner orange egg-shaped region in Fig. \ref{fig:GSosc_bound},
which is more tightly constrained than those of nucleon.

A closer inspection of Fig.~\ref{fig:EFFratio} shows that the data are not fully consistent with a constant ratio but may contain an energy dependent variation.
However, the precision of the data at present is insufficient for a finer exploration of possible oscillation structures \cite{Dai:2021yqr}. So $G_{\pm,0}^D$ and $G_{\pm,0}$ are not distinguished further.
As a consequence the information on $\phi_I$ can not be extracted from the relative phase for the $\Sigma^+$ and $\Sigma^-$ oscillation.
However, the measurement of $\Sigma^0$ channel supplies another constraint:
\bea \label{eq:Gsig0}
\frac{|G_0|^2}{|G_+|^2 + |G_-|^2} = \frac{1}{3 \delta_I^2 + 2 } &=&  0.27 \pm 0.04
\eea
depicted by the grey band in Fig. \ref{fig:EFFratio}.
Then the allowed region of $\delta_I$ and $\phi_I$ is the grass green area in Fig. \ref{fig:GSosc_bound}.
The value $\delta_I = 1$ is excluded with high statistical significance.
{The in-phase $\phi_I = 0$ is excluded with low significance. }
In the future these two methods could be used to cross-check the obtained constraint of the $\Sigma$ isospin amplitudes or quantify even the violation of isospin symmetry if more precise data of EFF are available.
Similar to Eq. (\ref{eq:ratioM}) and Eq. (\ref{eq:ratioE}), corresponding hyperon ratios $R_+$ and $R_-$ are extracted from the data of angular distribution as well.
{The $\Sigma^0$ channel further separates the moduli of isoscalar and isovector EMFFs:}
\be \label{eq:GsigM0}
|\frac{G^{0}_{M}}{\sqrt{3}}|^2 = (2\tau +1)  \frac{|G_0|^2}{2\tau + R_0^2} 
\ee
\be \label{eq:GsigM1}
|\frac{G^{1}_{M}}{\sqrt{2}}|^2 = \frac{2\tau +1}{2} \lf( \frac{|G_+|^2}{2\tau + R_+^2} + \frac{|G_-|^2}{2\tau + R_-^2} - \frac{2|G_0|^2}{2\tau + R_0^2} \rg) 
\ee
\be \label{eq:GsigE0}
|\frac{G^{0}_{E}}{\sqrt{3}}|^2 = (2\tau +1)  \frac{R_0^2|G_0|^2}{2\tau + R_0^2} 
\ee
\be \label{eq:GsigE1}
|\frac{G^{1}_{E}}{\sqrt{2}}|^2 = \frac{2\tau +1}{2} \lf( \frac{R_+^2|G_+|^2}{2\tau + R_+^2} + \frac{R_-^2|G_-|^2}{2\tau + R_-^2} - \frac{2 R_0^2|G_0|^2}{2\tau + R_0^2} \rg) 
\ee
The numerical calculation is postponed because of lacking of the data.
{A rough $R_{+}$ is measured by BESIII and statistics are not sufficient to extract $R_{-}$ and $R_{0}$ \cite{BESIII:2020uqk}. }

Assuming flavor symmetry { $G^{u/\Sigma^+}_{E,M} = G^{d/\Sigma^-}_{E,M} = G^{u/\Sigma^0}_{E,M} = G^{d/\Sigma^0}_{E,M} $,} the flavor decomposition of $\Sigma$ isotriplet is~\cite{Lin:2008mr}
\bea
G^+_{E,M} &=& - \frac{1}{3} G^s_{E,M} + \frac{2}{3} G^u_{E,M} \\
G^-_{E,M} &=& - \frac{1}{3} G^s_{E,M} -\frac{1}{3} G^d_{E,M} \\
G^0_{E,M} &=& - \frac{1}{3} G^s_{E,M}
\eea
The ratio in Eq.(\ref{eq:dataRIS}) means that light up and down quarks contribute significantly to the $\Sigma^\pm$ EMFFs {in low-$q^2$ region}.

A naive prediction from isoscalar diquark dominance is $G_{\Lambda} = - G_0$~\cite{Chernyak:1987nu}, consistent with the data but within big errors \cite{Zhou:2022jwr,BaBar:2007fsu,Ablikim:2021nba,BESIII:2021ccp,BESIII:2023ioy}, anticipating, however, a significant violation of that relation.
The ratios of $\Sigma^0$ to $\Lambda$ EFFs and cross sections deviate obviously from a constant over the covered energy range as indicated by these data.
The underlying reason, e.g. vector mesons above threshold \cite{Cao:2018kos} or a violation of $G_{\Lambda} = - G_0$, is not yet known.

\begin{figure}[t]
    \centering
    \includegraphics[width=\linewidth]{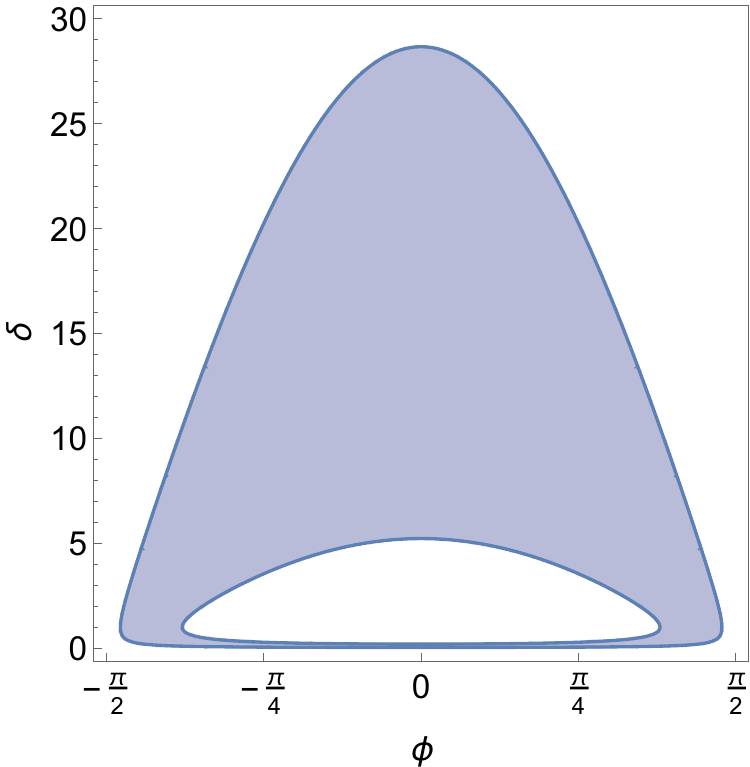}
    \caption{(Color online) Permeable boundaries of relative magnitude $\delta$ and phase $\phi$ for $\Xi$ baryon. }
    \label{fig:Gxi_bound}
\end{figure}

\paragraph{S=-2 Hyperons}
The cross section ratio between $e^{+}e^{-}\rightarrow\Xi^{0}\bar{\Xi}^{0}$ and $e^+e^-\to \Xi^-\bar{\Xi}^+$~\cite{BESIII:2020ktn,BESIII:2021aer,BESIII:2019cuv,Wang:2022zyc} from BESIII below charmonium region is 1.25 $\pm$ 0.15, consistent with 1.0 within 2$\sigma$.
The isospin decomposition of $\Xi^{-,0}$ is the same with nucleon, see Eq. (\ref{eq:isoamp}), resulting into
\be \label{eq:dataRIxi}
R_I^{\textrm{eff}} = \frac{|G^D_{\Xi^-}|^2 - |G^D_{\Xi^0}|^2}{|G^D_{\Xi^-}|^2 + |G^D_{\Xi^0}|^2}  = 0.22 \pm 0.15,
\ee
{This put a very loose constraint on the ratio of moduli $\delta$, as shown in Fig. \ref{fig:Gxi_bound}.
Comparing to Fig. \ref{fig:GSosc_bound}, the isospin components of nucleon, $\Sigma$ and $\Xi$ are explicitly distinguished by the present data.
The data of $R_{\Xi^0}$ and $R_{\Xi^-}$ are yet unavailable.}

Assuming flavor symmetry $G^{u/\Xi^-}_{E,M} = G^{d/\Xi^0}_{E,M}$, the quark flavor decomposition of $\Xi$ isospin doublet is~\cite{Lin:2008mr}
\bea
G^{\Xi^-}_{E,M} &=& - \frac{1}{3} G^s_{E,M} - \frac{1}{3} G^u_{E,M} \\
G^{\Xi^0}_{E,M} &=& - \frac{1}{3} G^s_{E,M} + \frac{2}{3} G^d_{E,M}
\eea
The nearly equivalent EFFs of $e^+e^-\to \Xi^-\bar{\Xi}^+$ and $\Xi^{0}\bar{\Xi}^{0}$ means that light up and down quarks contribute insignificant but the strange quarks are dominant in the EMFFs of $\Xi$ hyperons.

\section{About Charmonium decay} \label{sec:charm}

\begin{table*}[tbh]
\vspace{-2mm}
\centering
\caption{The isospin non-conservation percentage $R_I^{\textrm{eff}}$, $R_I^M$, and $R_I^E$ extracted from the branching ratios listed in Particle Data Group \cite{ParticleDataGroup:2022pth} and the measured $\alpha_B$ {(defined in Eq. (\ref{eq:alphaB}))} in literature.
}
\label{tab:charmonium}
\begin{tabular}{lccccc}
\hline\noalign{\smallskip}
Decay process & Branching ratio  & $\alpha_B$ & $R_I^{\textrm{eff}}$ & $R_I^M$ &  $R_I^E$\\ 
\noalign{\smallskip}\hline\noalign{\smallskip}
\hline
$J/\psi \to p \overline p$ & $(2.120 \pm 0.029) \times 10^{-3}$ & 0.595$\pm$0.012$\pm$0.015 \cite{BESIII:2012ion} & 0.007$\pm$0.039 & 0.02$\pm$0.05 & -0.11$\pm$0.26 \\ 
$J/\psi \to n \overline n$ & $(2.09 \pm 0.16) \times 10^{-3}$ & 0.50$\pm$0.04$\pm$0.21 \cite{BESIII:2012ion}
  & & & \\ 
\noalign{\smallskip}\hline\noalign{\smallskip}
$J/\psi \to \Xi^0 \overline \Xi^0$ & $(1.17 \pm 0.04) \times 10^{-3}$ & 0.66$\pm$0.03$\pm$0.05 \cite{BESIII:2016nix} & 0.09$\pm$0.04 & 0.10$\pm$0.04 & -0.01$\pm$0.10 \\ 
$J/\psi \to \Xi^- \overline \Xi^+$ & $(0.97 \pm 0.08) \times 10^{-3}$ &
0.586$\pm$0.012$\pm$0.010 \cite{BESIII:2021ypr} 
& & & \\ 
\noalign{\smallskip}\hline\noalign{\smallskip}
$J/\psi \to \Sigma^0 \overline \Sigma^0$ & $(1.172 \pm 0.032) \times 10^{-3}$ & -0.449$\pm$0.020$\pm$0.008 \cite{BESIII:2017kqw} & & & \\ 
$J/\psi \to \Sigma^+ \overline \Sigma^-$ & $(1.07 \pm 0.04) \times 10^{-3}$ & -0.508$\pm$0.006$\pm$0.004 \cite{BESIII:2020fqg} 
& & & \\ 
$J/\psi \to \Sigma^- \overline \Sigma^+$ &  --- & --- & & & \\
\noalign{\smallskip}\hline\noalign{\smallskip}
$J/\psi \to \Lambda \overline \Lambda$ & $(1.89 \pm 0.09) \times 10^{-3}$ &  0.461$\pm$0.006$\pm$0.007 \cite{BESIII:2018cnd} 
& & & \\ 
$J/\psi \to \Lambda \overline \Sigma^0 + {\rm c.c.} $ & $(2.83 \pm 0.23) \times 10^{-5}$ & --- & & & \\ 
\noalign{\smallskip}\hline\noalign{\smallskip}
\hline
$\psi(2S) \to p \overline p$ & $( 2.94 \pm 0.08) \times 10^{-4}$ & 1.03$\pm$0.06$\pm$0.03 \cite{BESIII:2018flj} & -0.020$\pm$0.028 & 0.02$\pm$0.04 & -1.0$\pm$0.8 \\
$\psi(2S) \to n \overline n$ & $(3.06 \pm 0.15) \times 10^{-4}$ & 0.68$\pm$0.12$\pm$0.11 \cite{BESIII:2018flj} & & & \\
\noalign{\smallskip}\hline\noalign{\smallskip}
$\psi(2S) \to \Xi^0 \overline \Xi^0$ & $(2.3 \pm 0.4) \times 10^{-4}$ & 0.65$\pm$0.09$\pm$0.14 \cite{BESIII:2016nix} & -0.11$\pm$0.08 & -0.12$\pm$0.08 & -0.04$\pm$0.28 \\
$\psi(2S) \to \Xi^- \overline \Xi^+$ & $(2.87 \pm 0.11) \times 10^{-4}$ & 0.693$\pm$0.048$\pm$0.049 \cite{BESIII:2022lsz} 
& & & \\ 
\noalign{\smallskip}\hline\noalign{\smallskip}
$\psi(2S) \to \Sigma^0 \overline \Sigma^0$ & $(2.35 \pm 0.09) \times 10^{-4}$ & 0.71$\pm$0.11$\pm$0.04 \cite{BESIII:2017kqw} & & & \\ 
$\psi(2S) \to \Sigma^+ \overline \Sigma^-$ & $(2.43 \pm 0.10) \times 10^{-4}$ & 0.682$\pm$0.03$\pm$0.011 \cite{BESIII:2020fqg} 
& -0.074$\pm$0.026 & -0.12$\pm$0.03 & 0.8$\pm$0.4 \\ %
$\psi(2S) \to \Sigma^- \overline \Sigma^+$ & $(2.82 \pm 0.09) \times 10^{-4}$ & 0.96$\pm$0.09$\pm$0.03 \cite{BESIII:2022ulr} & & & \\ 
\noalign{\smallskip}\hline\noalign{\smallskip}
$\psi(2S) \to \Lambda \overline \Lambda$ & $(3.81 \pm 0.13) \times 10^{-4}$ & 0.82$\pm$0.08$\pm$0.02 \cite{BESIII:2017kqw}  & & & \\
$\psi(2S) \to \Lambda \overline \Sigma^0 + {\rm c.c.} $ & $(1.6 \pm 0.7) \times 10^{-6}$ & --- & & & \\ 
\noalign{\smallskip}\hline\noalign{\smallskip}
\hline
\end{tabular}
\end{table*}%

The isospin analysis analyses discussed in the previous sections are also applicable to the decay of Charmonium, which is the topic of this section.
However, because of lack of data the discussion is of a more qualitative nature aiming at pointing out the perspectives of future experiments.
Considering the emergence of the more and more precise data, the results presented below may be helpful to prepare experiments and to interpret the upcoming experimental results~\cite{Wu:2021yfv}.
The Charmonium decay to $B \bar{B}$ is dominant by the isoscalar hadronic component with minor isospin violating a electromagnetic and a mixed strong-electromagnetic contributions \cite{Claudson:1981fj}.

The ratios $R_I^{\textrm{eff}}$, $R_I^M$, and $R_I^E$ serve as the quantification of isospin non-conservation to the total, electric and magnetic contribution.
As shown in Table \ref{tab:charmonium}, the branching ratios of $J/\psi\to B\bar{B}$ {(related to electromagnetic form factors through Eq. (\ref{eq:width}))} in the same isospin mutiplets are compatible within a relative accuracy of around 10\% \cite{BESIII:2012ion}, resulting into nearly zero $R_I^{\textrm{eff}}$, thus reflecting the approximate isospin symmetry.

If including the information on the measured $\alpha_B$ {(related to electromagnetic form factors through Eq. (\ref{eq:alphaB}))}, an important observation is that the isospin non-conservation mainly comes from the electric nucleonic decay amplitude and the magnetic $\Xi$ amplitude.
Remarkably, the isospin violating processes contribute potentially around 10\% for $J/\psi \to N\bar{N}$, indicated both by branching ratios and $\alpha_B$.
The isospin non-conservation from decay to $\Xi\bar \Xi$ contributes potentially around 10\%, attaining the same level between branching ratios and $\alpha_B$ of both $J/\psi$ and $\psi(2S)$.
The isospin violation part of electric amplitude of $\psi(2S) \to N \bar{N}$ and $\Sigma \bar{\Sigma}$ is of the same magnitude with the isospin conserving one. 
This is in sharp contrast with the level of isospin violation indicated in branching ratios, barely less than 10\%.
{Note that as shown by the errors of $R_I^E$ and $R_I^M$, the significance for the isospin violation contribution is only $2 \sigma$ at most, requesting for more accurate measurements.}

The data on $J/\psi \to \Sigma^0 \overline \Sigma^0$ and $J/\psi \to \Sigma^+ \overline \Sigma^-$ is already measured.
It is trivially to predict the values of $J/\psi \to \Sigma^- \overline \Sigma^+$ since they are equal to the other two channels under isospin conservation.
{The ratio of the branching ratios of $J/\psi \to \Sigma^0 \overline \Sigma^0$ and $J/\psi \to \Sigma^+ \overline \Sigma^-$ gives a very rough estimation for $\delta_I$:}
\be
0.022 < \delta_I < 1.4 
\ee
In analogy to the Eqs. (\ref{eq:GsigM0},\ref{eq:GsigE0}) the ratio $R_0 = 2.11 \pm 0.07$ of  $J/\psi \to \Sigma^0 \overline \Sigma^0$ channel is the relative modulus of isoscalar electric and magnetic amplitudes,
to be compared to $R_0 = 0.64 \pm 0.13$ extracted from the data of $\psi(2S) \to \Sigma^0 \overline \Sigma^0$ Channel.

It is feasible to explore the magnitude of isospin non-conservation if all decay channels are measured.
{This is accessible for $\psi(2S) \to \Sigma \overline \Sigma$ as seen in Table \ref{tab:charmonium} with the help of Eqs. (\ref{eq:GsigM0},\ref{eq:GsigM1},\ref{eq:GsigE0},\ref{eq:GsigE1}):
\be
\delta_I = 0.28 \pm 0.06, \phi_I = 88.1 \pm 0.7^\circ \mbox{or} -91.9 \pm 0.7^\circ \quad\\
\ee
\be
\delta_M = 0.33 \pm 0.07,  \phi_M = 78.2 \pm 1.0^\circ \mbox{or} -101.8 \pm 1.0^\circ \quad\\
\ee
\be
\delta_E = 0.48 \pm 0.23,  \phi_E = 65 \pm 15^\circ \mbox{or} -115 \pm 15^\circ
\ee
resulting into the same conclusion above that the isospin violation part of electric amplitude of $\Sigma \bar{\Sigma}$ is potentially larger, however, yet at the same level with magnetic and total amplitudes if considering the big errors.
This situation will probably be scrutinized in the future experiment considering the large error of $\alpha_B$ in $\psi(2S) \to \Sigma \bar{\Sigma}$.
It is also concluded that isospin violation and conservation parts tend to be orthogonal.}

The branching ratios of $J/\psi$ and $\psi(2S) \to \Lambda\bar{\Lambda}$ are found to be around 1.6 times larger than that of $\Sigma^0\bar{\Sigma}^0$, indicating a moderate breaking of isospin symmetry as expected.
The same level of violation is observed for $\psi$(3770) at CLEO-c Collaboration \cite{Dobbs:2014ifa}.
The branching ratio of isospin violation decay $J/\psi$ and $\psi(2S) \to \Lambda \overline \Sigma^0 + {\rm c.c.} $~\cite{BESIII:2012xdg,BESIII:2021mus}is around two orders in magnitude than that of the corresponding isospin allowed decays, hinting also to a small isospin violation in the decay width.

{If putting aside the big uncertainties in Table \ref{tab:charmonium},}
the scale of isospin violation is compatible with the analysis of SU(3) flavor symmetry~\cite{Mo:2021asa} and a parameterization of effective amplitudes~\cite{BaldiniFerroli:2019abd}.
It is in the same level as the background from the continuum, e.g. contributions of baryon structure in Sec. \ref{sec:nucleon}.
If higher precision of measurement is feasible in future,
it is definitely required to carefully examine the interference between charmonium and continuum amplitudes on the measurement of branching fractions and angular distributions \cite{Guo:2022gkg}.
For instance, it is anticipated that the the measured $J/\psi$ line shape is distorted from a Breit-Wigner form by this interference term in $ p\bar{p}$ and $n\bar{n}$ channels~\cite{Xia:2015mga,Mo:2021asa}.
{The angular distributions can be used to differential the origin of the isospin nonconservation from electro- and magnetic-coupling.
The polarization measurement of entangled hyperon pair produced in charmonium decay would provide more accurate data \cite{BESIII:2023lkg,BESIII:2023euh,BESIII:2021cvv,BESIII:2016ssr}.}
A precise calculation of $\alpha_B$ would surely be helpful \cite{Kivel:2021uzl,Kivel:2022fzk} if higher order corrections are properly considered \cite{Kivel:2022qjy,Kivel:2019wjh}.

\begin{figure}[t]
    \centering
    \includegraphics[width=\linewidth]{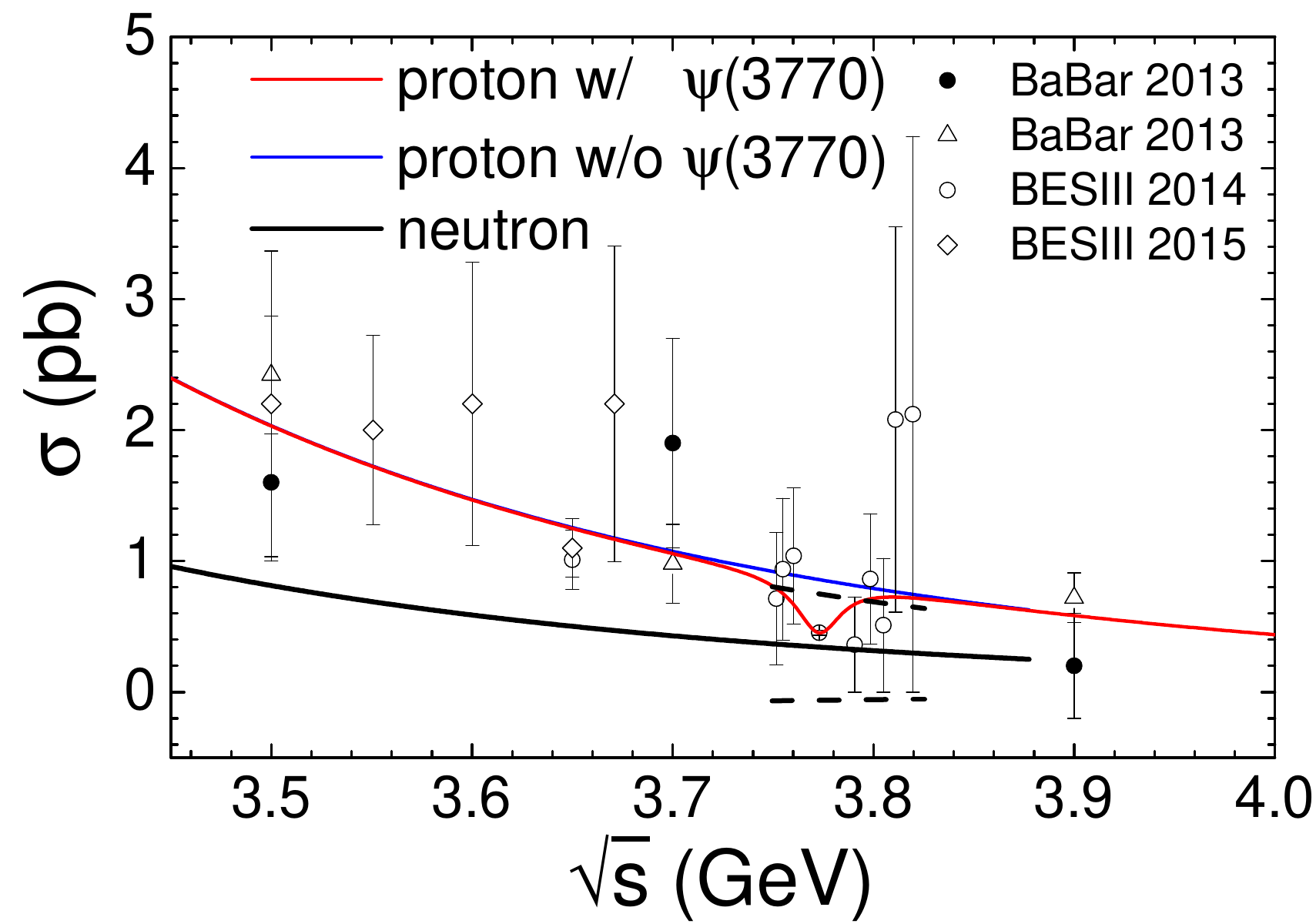}
    \caption{(Color online) The cross sections of $e^+ e^- \to p\bar{p}$ and $e^+ e^- \to n\bar{n}$ in the vicinity of $\psi(3770)$. The blue and black line are {the modified dipole} parameterization in Eq. (\ref{eq:dipole}).
    The red line is a coherent sum of Breit-Wigner form of $\psi(3770)$ with {modified} dipole form in Eq. (\ref{eq:dipole}).
    The dashed line is the predicted peak position of $\psi(3770)$ in $e^+ e^- \to n\bar{n}$.
    The data of $e^+ e^- \to p\bar{p}$ are from  BaBar \cite{BaBar:2013ukx,BaBar:2013ves} and BESIII \cite{BESIII:2014fwz,BESIII:2015axk}.}
    \label{fig:psipn}
\end{figure}

The precision of broad Charmonium states above $D\bar{D}$ threshold is far from being adequate for the investigation of isospin violation contribution.
In the near future more detailed measurement of the line shape of $\psi(3770)$ can be used to easily check our claim in Eq.~(\ref{eq:pexpansion}).
Fig. \ref{fig:psipn} shows the cross sections of $e^+ e^- \to p\bar{p}$ and $e^+ e^- \to n\bar{n}$ in the vicinity of ${\psi}(3770)$.
The dip in the $e^+ e^- \to p\bar{p}$ results from a destructive interference of ${\psi}(3770)$ and continuum.
Considering the unknown
$\Delta \phi = D_n-D_p$ in the charmonium region,
a solid prediction can only be made by Eq.~(\ref{eq:pexpansion}) to the range of peak or dip position of $\psi(3770)$ in $e^+ e^- \to n\bar{n}$, as indicated by the dashed line.
The upper line, e.g. a peak, is preferred for two reasons.
First, the lower line extends into nonphysical negative values.
Second, the $\Delta \phi$ is expected
to be weakly energy dependent and its value below 3.0 GeV favors a peak of $\psi(3770)$ in $e^+ e^- \to n\bar{n}$.
However, both reasons suffer from large errors so the dip shape of $\psi(3770)$ in $e^+ e^- \to n\bar{n}$ is not fully excluded.
{The analysis here is not intending to prove the existence or non-existence of a certain kind
meson resonances. Rather, they were referred to merely as auxiliary entities, serving as an attempt
to connect the derived phenomena to elements of hadron physics.}

As can be seen in Fig. \ref{fig:psipn}, the relative magnitude of $|I^{\textrm{BW}}_N/I^D_N|$ is around 0.3 for $\psi(3770)$ but around 0.1 $\sim$ 0.2 for light mesons, both of which is in the same level of relative accuracy of local data.
So the NLO contribution in Eq.~(\ref{eq:pexpansion}) can not be identified by the currently available data.
In other word, the data are well described both by the Breit-Wigner form and an oscillatory function, as demonstrated in detail in our previous work \cite{Cao:2021asd}.
Hopefully the precision of data in the vicinity of $\psi(3770) \to p\bar{p}$ will approach to the same level with that below 3.0 GeV and validate the NLO contribution.
Similar argument can be applied to $\psi(3770)\rightarrow\Lambda\bar\Lambda$ as well \cite{BESIII:2021ccp}.
However, this is an extremely trivial phenomenon, already known from Flatt\'{e} parametrizations or from a Fano line shape analysis of resonances, for instance, in the $\psi(3770) \to D\bar{D}$ reaction\cite{Wang:2014zha}.
The underlying mechanism includes three-gluon and $D$-meson loop contributions~\cite{Bystritskiy:2021frx,Bystritskiy:2022nqn,Qian:2021neg}.

\section{Summary and conclusion}

Based on the isospin analyses, the periodic oscillation structures of nucleon effective form factors in the timelike region are attributed to manifestation of light unflavored vector mesons.
The continuum part of proton and neutron EFFs is used to determine the range of the relative phase and {relative magnitude} between isoscalar and isovector component of EFFs.
We further point out that the relative phase is constrained by the phase difference between the oscillatory modulation of proton and neutron EFFs.
The isospin analyses are extended to the angular distributions.
Furthermore we successfully generalize the whole framework to the SU(3) octet baryons,
{and it is in fact applicable to decuplet baryons as well \cite{BESIII:2019dve,BESIII:2021gca}.}
An important result is that the range of the cross section of neutral $\Sigma$ hyperon channel can be predicted from the measured cross section of charged $\Sigma^\pm$ hyperon channel.

Isospin violation in the isovector form factors of the nucleon are predicted to be originated from
strange EMFFs as calculated by LQCD~\cite{Djukanovic:2019jtp} and by chiral perturbation theory~\cite{Kubis:2006cy}.
At present the timelike $N\bar{N}$ data are not precise enough to explore these kind of contribution.
Alternatively, we demonstrate that the precise data of $J/\psi$ and $\psi$(2S) decay allow for a first glimpse of the isospin non-conservation amplitudes within the same framework.

A better knowledge of the vector meson spectrum \cite{Wang:2021gle,Zhou:2022ark} will certainly improve our understanding of continuum part of the timelike nucleon EMFFs.
Their interfering contributions induce the oscillation structures seen in the cross section data.
A more dedicate study is required to go beyond the simple parameterization in Eq. (\ref{eq:osc}).
Though recently BESIII collaboration measured cross sections and invariant mass spectra of many channels with unprecedented accuracy, the properties of the vector mesons above $N \bar{N}$ threshold are still poorly understood.
The proper description of the interference between amplitudes resulting from the decay into wide vector mesons and continuum process is a demanding task for theory,
because e.g. a coupled-channel framework of multi-meson final states is by far too involved, if not impossible.
As a result, phenomenological fits are inadequate to extract precisely the masses and widths of vector mesons from data.
A complicated and time-consuming multi-channel partial wave analysis, possibly combined with a dispersion-theoretical framework, would be required to extract meson spectra for the available data.
Unfortunately, the meson spectra above $N \bar{N}$ threshold is hardly accessible in the constituent quark model \cite{Pang:2019ttv,Pang:2019ovr,Wang:2022xxi,Pang:2015eha,Wang:2021abg}.
Unquenched lattice QCD would be helpful but such calculations are yet unavailable.

In conclusion, even a full analysis of $e^+ e^-$ reactions does not allow by itself to determine the isospin of vector mesons, and input from theoretical side is indispensable.
At present it is premature to claim the isospin of the local structures in $e^+ e^- \to  N \bar{N}$, but instead, it is safely to identify them as vector mesons rather than something else, as clearly illustrated in our paper.
To bridge the seemly small gap between two conclusions, plenty of experimental and theoretical efforts are required, a disappointing fact that has to be accepted.
{We agree, of course, that the mentioned resonances are waiting for a solid confirmation.
However, whether they exist or not does not change the conclusions drawn from the isospin analysis alone.}
The isospin analyses in this paper give rise of another perspective providing that the isospin broken scale is
comparable to or smaller than the uncertainties of data.
With the help of available data, our analysis of isospin EFFs and EMFFs in the timelike region identified unambiguously the interference between them and unraveled the compatible range of parameters.
The conclusions are robust to large extent in view of its model independence.
So it paves the way for a fully consistent understanding of timelike and spacelike nucleon form factors.
{The proposed Super $\tau$-Charm facility (STCF)~\cite{Achasov:2023gey} gives the opportunity to further scrutinize the scenario discussed here.}


\bigskip

\begin{acknowledgments}

This work is supported by the National Natural Science Foundation of China (Grants Nos. U2032109, 1216050075, 12075289 and 12165022) and the Strategic Priority Research Program of Chinese Academy of Sciences (Grant NO. XDB34030301), and Yunnan Fundamental Research Project under Contract No. 202301AT070162.

\textbf{Data availability:}
Only public data was used in this research, see references in the text.

\end{acknowledgments}

\bigskip

\appendix

\section{Basics of $B\bar{B}$ Production and related Polarization Observables} \label{apx:polar}

The total cross section for $e^+e^- \to B\bar B$ reaction reads as:
\bea \label{eq:xsection}
\sigma_{B} &=& \frac{4\pi\alpha_{\textrm{em}}^2\beta C(q^2)}{3q^2}\lf[|G_M^B(q^2)|^2 + \frac{|G_E^B (q^2)|^2}{2\tau} \rg]
\eea
in terms of the Sachs electric and magnetic form factors $G_E$ and $G_M$. Here, $\alpha_{\textrm{em}}$ is the electromagnetic fine structure constant, $\beta$ is the velocity of the final state nucleon or anti-nucleon in the $e^+e^-$ certer of mass system, $\tau = q^2/4m_B^2$ with $s = q^2$ and $m_B$ being the invariant four-momentum transfer squared and the baryon mass, respectively, and $C(q^2)$ is the S-wave Sommerfeld-Gamow factor for the Coulomb correction \cite{Sakharov:1948plh}.
In order to avoid the complication from phase space and Coulomb factor, later on we mainly discuss the effective form factor (EFF):
 \bea \label{eq:eff}
|G^B_{\textrm{eff}}| &=& \sqrt{\frac{2\tau |G_M^B(q^2)|^2 + |G_E^B(q^2)|^2}{2\tau +1}}
\eea
which by definition could be extracted from the data of total cross section. The {electromagnetic form factor ratio}
\be \label{eq:RB}
R_B = \lf | \frac{G_E^B(q^2)}{G_M^B(q^2)} \rg |
\ee
could be determined by the data of different cross sections.
Usually the alternative variable $\alpha_B$ are given by experiment of charmonium decaying to $B\bar B$:
\be \label{eq:alphaB}
\alpha_B = \frac{\tau - R_B^2}{\tau + R_B^2}
\ee
with $\tau$ defined in the peak mass of the charmonium state.
{
The branching width is \cite{Kivel:2021uzl}
\be \label{eq:width}
\mcb (\psi \to B\bar B) = \frac{M_{\psi}\beta}{12 \pi \Gamma_{\psi}}\lf[|G_M^{\psi}|^2 + \frac{|G_E^{\psi}|^2}{2\tau} \rg]
\ee
}

If the produced antinucleon is polarized when unpolarized
electron beam is implemented,
or if one of the colliding beam is longitudinally polarized
then the antinucleon acquires $x$- and $z$-components of
the polarization, which lie in the reaction plane,
single-spin observables are defined as $P_y$ and $P_x$, respectively \cite{Denig:2012by}
\bea
P_y&=&\frac{\sin2\theta }{\sqrt{\tau }D} \mathrm{Re}(G_{M}G_{E}^*) \,\label{eq:Py}
\\ \label{eq:Px}
P_x&=&-\frac{2\sin\theta }{\sqrt{\tau }D}  \mathrm{Im}(G_{M}G_{E}^*)
\eea
with $D=(1+\cos^2\theta)(|G_{M}|^2)+\frac{1}{\tau } \sin^2\theta |G_{E}|^2 $.
There are no more independent polarization variables.

In terms of isospin decomposition of nucleon-anti-nucleon production amplitudes {which converts the basis from the physical states $(p, n)$ to isospin $(I_0, I_1)$}, one arrives at:
\bea
\Re \lf[ \frac{G^{u + d}_M}{3}  G^{u + d*}_E + G^{u - d}_M G^{u - d*}_E \rg ] &=& - \frac{\sqrt{\tau_p } D_p P_x^p + \sqrt{\tau_n } D_n P_x^n}{\sin\theta} \nn \\
\Re \lf[ \frac{G^{u + d}_M}{3}  G^{u - d*}_E + G^{u - d}_M G^{u + d*}_E \rg ] &=& - \frac{\sqrt{\tau_p } D_p P_x^p - \sqrt{\tau_n } D_n P_x^n}{\sin\theta}  \nn \\
\Im \lf[ \frac{G^{u + d}_M}{3}  G^{u + d*}_E + G^{u - d}_M G^{u - d*}_E \rg ] &=&  \frac{\sqrt{\tau_p } D_p P_y^p + \sqrt{\tau_n } D_n P_y^n}{\sin\theta \cos\theta} \nn \\
\Im \lf[ \frac{G^{u + d}_M}{3}  G^{u - d*}_E + G^{u - d}_M G^{u + d*}_E \rg ] &=& \frac{\sqrt{\tau_p } D_p P_y^p - \sqrt{\tau_n } D_n P_y^n}{\sin\theta \cos\theta}  \nn
\eea
which probe the relative magnitudes and phases between $G^{u \pm d}_M$ and $G^{u \pm d}_E$.

In conclusion, four independent complex variables (2 isospin $\times$ 2 electromagnetic form factors), e.g. seven independent real variables considering only relative phases are relevant, can be extracted from
eight independent observables (total and differential cross sections, two polarization observables for proton and neutron) in each energy point.
{Similar conclusion is arrived in the basis of physical states $(p, n)$.}

\bibliography{strangenote}

\end{document}